\documentclass{article}
\usepackage{graphicx} 
\usepackage{geometry}  
\usepackage{amsmath}
\title{Contrasting the Implicit Method in Incoherent Lagrangian and the Correction Map Method in Hamiltonian}
\author{Junjie Luo$^{1,}$*$^{,\dagger}$, Jie Feng $^{2,}$*$^{,\dagger}$, Hong-Hao Zhang$^{1,}$*$^{,\dagger}$ and Weipeng Lin $^{3,}$*$^{,\dagger}$}
\date{August 2023}
\makeatletter 
\begin{document}
\maketitle

DOI:https://doi.org/10.3390/sym15071401

Pubblished by Symmetry\\
$^{1}$ \quad School of Physics, Sun Yat-sen University, Guangzhou 510275, China\\
$^{2}$ \quad School of Science, Sun Yat-sen University, Shenzhen 518107,  China\\
$^{3}$ \quad School of Physics and Astronomy, Sun Yat-sen University, Zhuhai 519000,  China

\abstract{The equations of motion for a Lagrangian mainly refer to the acceleration equations, which can be obtained by the Euler--Lagrange equations. In the post-Newtonian Lagrangian form of general relativity, the Lagrangian systems can only maintain a certain post-Newtonian order and are incoherent Lagrangians {since the higher-order terms are omitted}. This truncation can {cause some changes in the constant of motion}. However, in celestial mechanics, Hamiltonians are more commonly used than Lagrangians. {The conversion from Lagrangian
 to Hamiltonian} can be achieved through the Legendre transformation. The coordinate momentum separable Hamiltonian can be computed by the symplectic algorithm, whereas the inseparable Hamiltonian can be used to compute the evolution of motion by the {phase-space expansion} method. Our recent work involves the design of a multi-factor correction map for the phase-space expansion method, known as the correction map method. In this paper, we compare the performance of the implicit algorithm in post-Newtonian Lagrangians and the correction map method in post-Newtonian Hamiltonians. Specifically, we investigate the extent to which both methods can {uphold invariance of the motion's constants, such as energy conservation and angular momentum preservation}. Ultimately, the results of numerical simulations demonstrate the superior performance of the correction map method, particularly with respect to angular momentum conservation.}

\section{Introduction}
Compact binary systems, composed of neutron stars or black holes, etc., are of immense interest to experimental and theoretical researchers as sources of gravitational waves for broadband laser interferometers. {The temporal progression of binary systems encompassing compact objects can be elucidated through the implementation of Einstein's equations of general relativity.} Explicit symplectic integrators are supposed to be the ideal candidate with several benefits for the numerical simulations in these systems. They are designed to preserve the symplectic structure, which {guarantees the precision and stability of numerical solutions} over long time intervals. However, 
Einstein's equations of general relativity describe the motion of strong gravitational systems for which exact solutions are very difficult to obtain, but there have been some efforts to address this issue. {For example, Xin Wu 
 \cite{rr19} developed the explicit symplectic integrators for Hamiltonian systems in curved spacetimes, particularly for black hole spacetimes. The papers \cite{rr20, rr18} discuss the motion of charged particles around the Schwarzschild black hole with an external magnetic field. Therefore, the orbital motion described in this paper is a two-body problem, involving the motion of a charged particle around a Schwarzschild black hole. Ying wang \cite{rr21, rr22} constructs the explicit symplectic integrators in general relativity, specifically for the Hamiltonian of Schwarzschild spacetime geometry. The integrators are useful for the long-term integration of N-body Hamiltonian systems and modeling the chaotic motion of charged particles around a black hole with an external magnetic field. Xin Wu \cite{rr23} discusses the construction of explicit symplectic integrators for Kerr black holes in general relativity. The authors introduce a time transformation function to the Hamiltonian of Kerr geometry to obtain a time-transformed Hamiltonian consisting of five splitting parts whose analytical solutions are explicit functions of the new coordinate time. Wei Sun \cite{rr24} proposes an explicit symplectic integrator for the Kerr spacetime geometry to simulate the nonintegrable dynamics of charged particles moving around the Kerr black hole in an external magnetic field. The algorithm shows good numerical performance and is used to study the dynamics of order and chaos of charged particles.}

Despite the efficacy of this method, another useful and well-developed approach is the form of the post-Newtonian (PN) Lagrangian or Hamiltonian \cite{rr25} \mbox{approximation \cite{r4, r5, rr25}}. {Arun et al. \cite{rr30} investigated inspiralling compact binaries in quasi-elliptical orbits and provided a comprehensive analysis of the third post-Newtonian energy flux. Their study focused on understanding the energy loss due to gravitational radiation and its implications for compact binary systems. Tessmer and Schäfer \cite{rr31} studied the eccentric motion of spinning compact binaries. They examined the dynamics of these systems with non-circular orbits, considering the effects of spin and exploring the consequences of eccentricity on the gravitational wave signals emitted during inspiral. Hinder et al. \cite{rr32} developed an eccentric binary black hole waveform model by combining numerical relativity simulations with post-Newtonian theory. Their work aimed to accurately describe the complete inspiral--merger--ringdown phase of eccentric binary black hole systems, providing insights into the gravitational waveforms emitted during these events. Chattaraj et al. \cite{rr33} conducted high-accuracy comparisons between post-Newtonian theory and numerical relativity simulations, specifically focusing on eccentric binary black holes. They investigated the influence of higher modes on the waveforms and developed a model that incorporates eccentricity and accurately describes the inspiral, merger, and ringdown phases. Chowdhury and Khlopov \cite{rr34} studied an eccentric binary black hole system within the framework of post-Newtonian theory. Their research aimed to understand the behavior of binary black holes with non-circular orbits, providing insights into the dynamics and gravitational wave emissions of eccentric binary systems.} These approximations provide high-precision theoretical templates of gravitational waveforms, although their higher-order terms are truncated, which affects their equivalence \cite{r1, r2, r3}.

The choice of approximation and the selection of the algorithm becomes crucial to ensuring {an accurate and effective description of the trajectory evolution of compact binaries systems} and matching corresponding gravitational waveforms. 

The PN Lagrangian equations of motion are derived from the Euler--Lagrangian equations of a PN Lagrangian formulation, {denoted as $L(\mathbf{r},\mathbf{v})$. By calculating the partial derivative of $L$ with respect to velocity, we obtain the generalized momentum $\mathbf{p}=\partial L/\partial \mathbf{v}$. Similarly, the acceleration equations $\mathbf{a}=f(\mathbf{r},\mathbf{v},\mathbf{a}) =\partial L/\partial \mathbf{r}$ can be derived} and we can obtain a coherent Lagrangian \cite{r6, rr28, rr29}. {By limiting the inclusion of accelerations up to a certain PN order in the Lagrangian, the accelerations $\mathbf{a}$ in the function $f$ will be modified to $\mathbf{a}^*$; it only has lower-order terms, i.e., $\mathbf{a}=f(\mathbf{r},\mathbf{v},\mathbf{a}^*)$.}
 The acceleration equations become incoherent, due to the higher-order PN terms disappearing, leading to a loss of some values of the constants of motion during subsequent evolution. The same problem occurs with the post-Newtonian Hamiltonian form. The error of the constant of motion can be used as an indicator to test the performance of different algorithms in both approximate forms.

Various algorithms are available for the calculation of post-Newtonian Lagrangian quantities. For example, {in optimizing the fifth-order Runge--Kutta method as a high-precision integrator, Zhong \cite{r7} employed corrections to all integrals within the conservative 3PN order Hamiltonian. Tsang \cite{r8} introduced an implicit symplectic integrator that accounts for 2.5PN 
 gravitational radiation reaction terms in the Newtonian two-body problem. This approach effectively captures the effects of radiation reactions. Lubich \cite{r9} devised an explicit and implicit mixed symplectic integration technique that facilitates the splitting of orbital and spin contributions. By employing this approach, the dynamics of both orbital and spin variables can be accurately simulated. Zhong \cite{r10} proposed fourth-order canonical explicit and implicit mixed symplectic methods. These methods offer improved accuracy and stability in the computation of post-Newtonian quantities. \mbox{Seyrich \cite{r11}} developed Gauss Runge--Kutta implicit canonical symplectic schemes that preserve the structural properties of the system. These schemes ensure long-term numerical stability and accuracy. These algorithms, with their distinct methodologies, contribute to advancing the computation of post-Newtonian Lagrangian quantities, addressing specific aspects such as precision, radiation reaction, spin contributions, stability, and \mbox{structural preservation}.}

Regarding the post-Newtonian Hamiltonian, the {phase-space expansion} \mbox{method \cite{r12, rr26, rr27}} is a usable algorithm. {The Hamiltonian lacks separability and does not possess a coordinate momentum or multiple integrable splitting components. Pihajoki \cite{r12} extended the phase space variables by copying the coordinates and momenta. We achieved a Hamiltonian splitting form so that the explicit leapfrog algorithms become available. The permutation map of momentum} was designed to suppress the interaction of the original and extended variables. {Liu \cite{r13} devised a sequential mapping of coordinate and momentum permutations} and constructed fourth-order {phase-space expansion} explicit method compositions of two triple products of the usual second-order leapfrog. These algorithms suffer a clear failure when calculating the chaotic orbits of celestial systems. The interactions between the original {variables and the extended one} become increasingly strong and show considerably different values, whereas they are supposed to be equivalent. Midpoint and correction maps \cite{r14, r15} have been proposed to ensure the equivalence of the {original variables and the copy one}. Recently, we proposed a multi-factor correction map that yields a higher accuracy of the phase-space expansion method without significant computational resource increases \cite{r16}. This paper aims to design a multi-factor correction map for post-Newton Hamiltonian and examine its performance.

This article is divided into several sections. {In Section \ref{sec:2}, we revisit the Lagrangian and Hamiltonian equations of motion for compact binary systems within the post-Newtonian (PN) approximation. Section \ref{sec:3} presents the introduction of the {phase-space expansion} method and the development of a correction map for the post-Newtonian Hamiltonian. In Section \ref{sec:4}, we conduct a comparative analysis of the accuracy of numerical solutions obtained using the implicit midpoint method in the computation of the post-Newtonian Lagrangian} and the correction map method for the post-Newtonian Hamiltonian. Finally, in Section \ref{sec:5}, we conclude
\section{PN Lagrangian and Hamiltonian in Compact Binary}\label{sec:2}
Let us consider a compact binary system governed by a PN Lagrangian $L(\mathbf{r}, \mathbf{v})$ up to the $m$-th order, where $\mathbf{r}$ and $\mathbf{v}$ represent the position and velocity vectors, respectively. The Euler--Lagrangian equation is given by Equation \ref{eq:1}, where the generalized momentum $\mathbf{p}$ is defined by Equation \ref{eq:2}. The expression for $\mathbf{p}$ is given by Equation \ref{eq:2} and represents a nonlinear algebraic equation of $\mathbf{v}$.

\begin{equation}
\frac{d\mathbf{p}}{dt}=\frac{\partial L}{\partial \textbf{r}}.\label{eq:1}
\end{equation}

Here 

\begin{equation}
\mathbf{p}=\frac{\partial L}{\partial \mathbf{v}};\label{eq:2}
\end{equation}

\begin{equation}
\frac{d\mathbf{r}}{dt}=\mathbf{v}.\label{eq:3}
\end{equation}

 We 
 note that Equations \ref{eq:2} and \ref{eq:3} are differential equations, and that $(\mathbf{r}, \mathbf{p})$ are treated as the integration variables, whereas $\mathbf{v}$ is not. However, we can substitute Equation \ref{eq:2} into Equation \ref{eq:3} to obtain the corresponding acceleration equation, given by Equation \ref{eq:4}. Here, $\mathbf{a}_{N},\mathbf{a}_{1PN},\mathbf{a}_{2PN},\dots,\mathbf{a}_{mPN}$ {correspond to the Newtonian term, the 1st, 2nd post-Newtonian-order term to the $m$-th post-Newtonian-order contributions for the accelerations
}.

\begin{equation}
\frac{d\mathbf{v}}{dt}=\mathbf{a}_{N}+\mathbf{a}_{1PN}+\mathbf{a}_{2PN}+......\mathbf{a}_{mPN}.\label{eq:4}
\end{equation}

When considering only the $m$-th PN order term in Equation \ref{eq:4}, all terms higher than the $m$-th PN order are truncated. {Consequently, Equation \ref{eq:4} does not align with the PN Lagrangian $L$, and Equations \ref{eq:3} and \ref{eq:4} are treated as incoherent PN equations of motion in the Lagrangian $L$. 
However, when utilizing Equations \ref{eq:3} and \ref{eq:4}, the variables $(\mathbf{r},\mathbf{v})$ can be used as a set of integration variables instead of the variables $(\mathbf{r},\mathbf{p})$. Nevertheless, this approach does not fully maintain constants of motion, such as the energy integral expressed by}

\begin{equation}
E=\mathbf{v}\cdot\mathbf{p}-L.\label{eq:5}
\end{equation}

{In this paper, $L$ is} the dimensionless post-Newtonian (PN) Lagrangian formulation {for compact binaries}. The evolution of binaries can be given by the expression:

\begin{equation}
L=L_{N}+L_{1PN}. \label{L}
\end{equation}

$L_N$ and $L_{1PN}$ denote the non-relativistic and 1PN contributions to the Lagrangian, respectively. {For simplicity, higher-order terms are not considered.} The non-relativistic part is expressed as:

\begin{equation}
L_{N}=\frac{\mathbf{\dot{r}}^{2}}{2}+\frac{1}{r}, 
\end{equation}

whereas the 1PN part is given by \cite{r4}:

\begin{equation}
L_{1PN} = \frac{1}{c^2}\left\{\frac{1}{8}(1-3\eta)v^{4}+\frac{1}{2r}[(3+\eta)v^{2}
+\frac{\eta}{r^2}(\mathbf{r}\cdot\mathbf{v})^{2} -\frac{1}{r} ]\right\}.
\end{equation}

{Here $\eta=\mu/M$ is the dimensionless mass parameter. The reduced mass, $\mu$, is defined as $M_1M_2/M = \beta(1+\beta)^{-2}$, and $\beta=M_1/M_2$ is the mass ratio, where $M_1$ and $M_2$ represent the masses of the two bodies constituting the binary system and the total mass is denoted as $M = M_1 + M_2$.  Additionally, $c$ is the speed of light and $G$ represents the constant of gravity given in natural units with $c=G=1$. $c$ is retained in some of the latter equations, and it can be ignored in the actual calculation.}

The equations for the evolution of the system can be derived from the Lagrangian formulation. According to Equations \ref{eq:1} and \ref{eq:2}, the equation for the evolution of the momentum can be written as:

\begin{equation}
\frac{d\mathbf{p}}{dt}=-\frac{\mathbf{r}}{r^3} \left\{ 1+\frac{1}{c^2}[\frac{3\eta}{2r^2}(\mathbf{r}\cdot\mathbf{v})^2+\frac{3+\eta}{2}v^2-\frac{1}{r}] \right\}+\frac{\eta}{c^2r^3}(\mathbf{r}\cdot\mathbf{v})\mathbf{v},
\end{equation}

where $\mathbf{r}$ is the separation vector between the two masses. The non-relativistic and 1PN contributions to this equation are, respectively, expressed in the first and second terms.
The expression for the generalized momentum $\mathbf{p}$ {till 1pN} is given in terms of the velocity $\mathbf{v}$ as:

\begin{equation}
\mathbf{p}=\mathbf{v}+\frac{1}{c^2}\left\{ \frac{v^2}{2}(1-3\eta)\mathbf{v}+\frac{1}{r}[\frac{\eta}{r^2}(\mathbf{r}\cdot\mathbf{v})\mathbf{r}+(3+\eta)\mathbf{v}]\right\}. \label{eq:11}
\end{equation}

The
 value of momentum can be obtained from Equation \ref{eq:11} once the velocity is known and vice versa. However, velocity needs to be solved iteratively since it cannot be obtained directly from the Lagrangian. The first post-Newtonian relative acceleration equation is used to determine the value of velocity, given by:

\begin{equation}
\frac{d\mathbf{v}}{dt}=\mathbf{a}_{N}+\mathbf{a}_{1PN}.\label{eq:12}
\end{equation}

The
 sub-terms are

\begin{equation}
\mathbf{a}_{N}=-\frac{\mathbf{r}}{r^3},
\end{equation}

\begin{equation}
\mathbf{a}_{1PN}=-\frac{1}{r^2c^2}\left\{  \frac{\mathbf{r}}{r}[(1+3\eta)v^2-\frac{2}{r}(2+\eta)-\frac{3\eta}{2r^2}(\mathbf{r}\cdot\mathbf{v})^2]-\frac{2}{r}(2-\eta)(\mathbf{r}\cdot\mathbf{v})\mathbf{v} \right\},
\end{equation}

Here, $\mathbf{a}_{N}$ and $\mathbf{a}_{1PN}$ describe the non-relativistic and 1PN contributions to the acceleration.
{Using Equations \ref{L} and   \ref{eq:11},} the energy integral in Equation \ref{eq:5} can be expressed as. 

\begin{equation}
E=\frac{v^2}{2}-\frac{1}{r}+\frac{1}{c^2}\left\{ \frac{3}{8}(1-3\eta)v^4+\frac{1}{2r}[(3+\eta)v^2+\frac{\eta}{r^2}(\mathbf{r}\cdot\mathbf{v})^2+\frac{1}{r}]. \right\}\label{eq:6}
\end{equation}

In summary, the Lagrangian formulation of the evolution of binaries provides a mathematical framework for studying their motion. The momentum and velocity of the system can be determined from the equations derived from the Lagrangian, which include non-relativistic and relativistic contributions to the acceleration. The dimensionless PN Lagrangian {offers valuable insights for the dynamics system}, enabling the study of gravitational wave emission caused by binary systems.

With Equations \ref{eq:3} and \ref{eq:12}, we can obtain the numerical solution $(\mathbf{r}\cdot\mathbf{v})$ by using the fourth-order implicit midpoint method ($IM_4$).

PN Hamiltonian form $H$ can be {derived through Lagrangian $L$ using the Legendre transformation},

\begin{equation}
H=\mathbf{p}\cdot \mathbf{\dot{r}}-L.
\end{equation}

Then we obtain the 1PN Hamiltonian,

\begin{equation}
H=H_{N}+H_{1PN}. \label{eq:16}
\end{equation}

In order to compare the effect of higher-order PN terms on the error in the constants of motion, we introduce the 2PN post-Newton Hamiltonian,

\begin{equation}
H^*=H_{N}+H_{1PN}+H_{2PN}. \label{eq:17}
\end{equation}

The expressions for the sub-terms in Hamiltonians \ref{eq:16} and \ref{eq:17} are, respectively, given by
 \begin{eqnarray}
H_{N}=T(\textbf{p})+V(\textbf{r})=\frac{\textbf{p}^{2}}{2}-\frac{1}{r},
\end{eqnarray}
\vspace{-6pt}
\begin{eqnarray}
H_{1PN} = \frac{1}{8}(3\eta-1)\textbf{p}^{4}-\frac{1}{2}[(3+\eta)\textbf{p}^{2} \nonumber \\
 +\frac{\eta}{r}(\textbf{r}\cdot\textbf{p})^{2}]\frac{1}{r}+\frac{1}{2r^{2}},
\end{eqnarray}
\vspace{-6pt}
\begin{eqnarray}
H_{2PN} &=& \frac{1}{16}(1-5\eta+5\eta^{2})\textbf{p}^{6}+\frac{1}{8}[(5-20\eta-
3\eta^{2})\textbf{p}^{4} \nonumber\\ && -\frac{2\eta^2}{r}(\textbf{r}\cdot\textbf{p})^{2}\textbf{p}^{2}-\frac{3\eta^2}{r}
(\textbf{r}\cdot\textbf{p})^{4}]\frac{1}{r} \nonumber\\
&& +\frac{1}{2}[(5+8\eta)\textbf{p}^{2}+\frac{3\eta}{r}
(\textbf{r}\cdot\textbf{p})^{2}]\frac{1}{r^{2}} \nonumber\\
&& -\frac{1}{4}(1+3\eta)\frac{1}{r^{3}},
\end{eqnarray}
Due to the disappearance of higher-order terms, $H$ and $H^*$ are approximately equal to $E$, and not strictly equivalent. The integrators used in the Hamiltonian $H$ and $H^*$ will be described in the next section.
\section{{Phase-Space Expansion} Method with a Multi-Factors Correction Map}\label{sec:3}
Since neither the Hamiltonian $H$ nor $H^*$ {can be separated into multiple integrable parts, the symplectic leapfrog method cannot be applied directly to these Hamiltonians} unless they are suitably modified to a splitting form. An effective approach to solving this problem is the {phase-space expansion} method. Pihajoki~\cite{r12} introduced a new pair of canonical and conjugate variables ($\widetilde{\textbf{r}}$, $\widetilde{\textbf{p}}$) from the original variables $(\mathbf{r}, \mathbf{p})$. This doubles the phase-space variables, $(\mathbf{r}, \mathbf{\widetilde{r}}, \mathbf{p}, \mathbf{\widetilde{p}})$ and constructs a new Hamiltonian $\widetilde{H}(\mathbf{r}, \mathbf{\widetilde{r}}, \mathbf{p}, \mathbf{\widetilde{p}})$ using two identical Hamiltonians $H_1$ and $H_2$:
\begin{eqnarray}
       \widetilde{H}(\textbf{r},\widetilde{\textbf{r}},\textbf{p},
       \widetilde{\textbf{p}})=H_{1}(\textbf{r},
       \widetilde{\textbf{p}})+H_{2}(\widetilde{\textbf{r}},\textbf{p}).
\end{eqnarray}
where both $H_1$ and $H_2$ should be equal to the original Hamiltonian $H$. The new Hamiltonian $\widetilde{H}$ already{ exhibits two integrable components. A conventional second-order leapfrog algorithm can be employed for its integration}:

\begin{equation}
\mathbf{A}_2(h)=\textbf{H}_{2}(\frac{h}{2})\textbf{H}_{1}(h)\textbf{H}_{2}(\frac{h}{2}), \label{A2}
\end{equation}

where $h$ represents the time step, $\textbf{H}_{1}$ and $\textbf{H}_{2}$ are Hamiltonian operators. {The code corresponding to integration \ref{A2} from $n$th to $(n+1)$th step is}
\begin{align}
&\textbf{r}_{n+\frac{1}{2}}=\textbf{r}_{n}+\frac{h}{2}\nabla_{\textbf{p}}H_{2}(\widetilde{\textbf{r}}_{n},\textbf{p}_{n})\nonumber\\
&\widetilde{\textbf{p}}_{n+\frac{1}{2}}=\widetilde{\textbf{p}}_{n}-\frac{h}{2}\nabla_{\widetilde{\textbf{r}}}H_{2}(\widetilde{\textbf{r}}_{n},\textbf{p}_{n})\nonumber\\
&\widetilde{\textbf{r}}_{n+1}=\widetilde{\textbf{r}}_{n}+h\nabla_{\widetilde{\textbf{p}}}H_{1}(\textbf{r}_{n+\frac{1}{2}},\widetilde{\textbf{p}}_{n+\frac{1}{2}})\nonumber\\
&\textbf{p}_{n+1}=\textbf{p}_{n}-h\nabla_{\textbf{r}}H_{1}(\textbf{r}_{n+\frac{1}{2}},\widetilde{\textbf{p}}_{n+\frac{1}{2}})\nonumber\\
&\textbf{r}_{n+1}=\textbf{r}_{n+\frac{1}{2}}+\frac{h}{2}\nabla_{\textbf{p}}H_{2}(\widetilde{\textbf{r}}_{n+1},\textbf{p}_{n+1})\nonumber\\
&\widetilde{\textbf{p}}_{n+1}=\widetilde{\textbf{p}}_{n+\frac{1}{2}}-\frac{h}{2}\nabla_{\widetilde{\textbf{r}}}H_{2}(\widetilde{\textbf{r}}_{n+1},\textbf{p}_{n+1}).
\end{align}
It can be seen that the calculation of the numerical solution $(\mathbf{r}, \mathbf{\widetilde{p}})$ of $H_2$ requires the numerical solution $(\widetilde{\textbf{r}},\textbf{p})$ of $H_1$, and vice versa, so there is an energy exchange between $H_1$ and $H_2$, and even if the initial conditions are the same, $H_1$ and $H_2$ will become unequal in the later evolution unless the errors are constant equal to 0. To submit the accuracy, we construct a fourth-order algorithm using Yoshida’s triplet product

\begin{equation}
\textbf{A}_4(h)=\textbf{A}_{2}(\lambda_{3} h)\textbf{A}_{2}(\lambda_2 h)\textbf{A}_{2}(\lambda_1 h).
\end{equation}

The time coefficients $\lambda_{1}$, $\lambda_{2}$, and $\lambda_{3}$ are identical to those presented in \cite{r17} and are set to $\lambda_{1}=\lambda_{3}=1/(2-2^{1/3})$ and $\lambda_{2}=1-2\lambda_{1}$. Algorithm $\textbf{A}_4$ is utilized to obtain a set of numerical solutions $(\mathbf{r}, \widetilde{\textbf{r}}, \mathbf{p}, \widetilde{\textbf{p}})$.
It is crucial to note that the original variables $(\mathbf{r}, \mathbf{p})$ and their {counterparts ($\widetilde{\textbf{r}}$, $\widetilde{\textbf{p}}$) are intended to be identical at each integration step, but in reality, they exhibit discrepancies. The interaction between the solutions $(\textbf{r},\widetilde{\textbf{p}})$ of $H_{1}$ and $(\widetilde{\textbf{r}},\textbf{p})$ of $H_{2}$ leads to their divergence over time}. To ensure the equivalence of the original {variables and their copy one}, Pihajoki \cite{r12} proposed a momentum permutation map, whereas Liu \cite{r13} proposed coordinate and momentum permutation maps. These applications were successful in several examples, but not in chaotic orbits. However, our previous \mbox{work \cite{r15, r16}} proposed the manifold corrections map, which effectively overcame the challenge.
Unlike that in the original paper \cite{r15, r16}, the correction map for the 1PN Hamiltonian is
\begin{eqnarray}
\textbf{M}_{1PN}=\left(\begin{array}{cccc}
\frac{\gamma}{2},  \frac{\gamma}{2},  \textbf{0},  \textbf{0} \\
\frac{\gamma}{2}, \frac{\gamma}{2},  \textbf{0},  \textbf{0} \\
\textbf{0},  \textbf{0}, \frac{\alpha}{2},  \frac{\alpha}{2} \\
\textbf{0},  \textbf{0},   \frac{\alpha}{2},  \frac{\alpha}{2}
\end{array}\right).
\end{eqnarray}
Here, {the momentum scaling factor $\gamma$ and the coordinate scaling factor $\alpha$ are incorporated into $\textbf{M}_{1PN}$ and can be obtained by solving the following formulations:}
\begin{align}
       T(\frac{\alpha \textbf{p}+\alpha \widetilde{\textbf{p}}}{2})=\frac{\widetilde{T}(\textbf{p},\widetilde{\textbf{p}})}{2} =\frac{T_{1}(\widetilde{\textbf{p}})+T_{2}(\textbf{p})}{2}, \label{eq:HT}
\end{align}
\begin{align}
      & V(\frac{\gamma\textbf{r}+\gamma\widetilde{\textbf{r}}}{2})+H_{1PN}(\frac{\gamma\textbf{r}+\gamma\widetilde{\textbf{r}}}{2},\frac{\alpha\textbf{p}+\alpha\widetilde{\textbf{p}}}{2}) \nonumber\\
      & =\frac{\widetilde{V}(\textbf{r},\widetilde{\textbf{r}})+\widetilde{H}_{1PN}(\textbf{r},\widetilde{\textbf{r}},\textbf{p},\widetilde{\textbf{p}})}{2}. \label{eq:HPN}
\end{align}
Equation \ref{eq:HT} provides $\alpha=\sqrt{\frac{2(\textbf{p}^2+\widetilde{\textbf{p}}^2)}{(\textbf{p}+\widetilde{\textbf{p}})^2}}$, and Newton's method is used to obtain $\gamma$ from Equation \ref{eq:HPN}. Then, the fourth-order {phase-space expansion} method with multi-factor correction map for 1PN Hamiltonian $H$ is established as

\begin{equation}
\textbf{CM}_{1PN}(h)=\textbf{M}_{1PN}\otimes\textbf{A}_4(h).
\end{equation}

Similarly, for the 2PN Hamiltonian $H^*$, the aforementioned steps are applicable. The correction map $\textbf{M}_{2PN}$ for the Hamiltonian $H^*$ follows the same structure as $\textbf{M}_{1PN}$.
\begin{eqnarray}
\textbf{M}_{2PN}=\left(\begin{array}{cccc}
\frac{\gamma}{2},  \frac{\gamma}{2},  \textbf{0},  \textbf{0} \\
\frac{\gamma}{2}, \frac{\gamma}{2},  \textbf{0},  \textbf{0} \\
\textbf{0},  \textbf{0}, \frac{\alpha}{2},  \frac{\alpha}{2} \\
\textbf{0},  \textbf{0},   \frac{\alpha}{2},  \frac{\alpha}{2}
\end{array}\right).
\end{eqnarray}
However, the solution for $\gamma$ is replaced by the following equation
\begin{align}
      & V(\frac{\gamma\textbf{r}+\gamma\widetilde{\textbf{r}}}{2})+H_{1PN}(\frac{\gamma\textbf{r}+\gamma\widetilde{\textbf{r}}}{2},\frac{\alpha\textbf{p}+\alpha\widetilde{\textbf{p}}}{2})+H_{2PN}(\frac{\gamma\textbf{r}+\gamma\widetilde{\textbf{r}}}{2},\frac{\alpha\textbf{p}+\alpha\widetilde{\textbf{p}}}{2}) \nonumber\\
      & =\frac{\widetilde{V}(\textbf{r},\widetilde{\textbf{r}})+\widetilde{H}_{1PN}(\textbf{r},\widetilde{\textbf{r}},\textbf{p},\widetilde{\textbf{p}})+\widetilde{H}_{2PN}(\textbf{r},\widetilde{\textbf{r}},\textbf{p},\widetilde{\textbf{p}})}{2}.
\end{align}
For convenience, such algorithms are referred to as the correction map method. The correction map method for  2PN Hamiltonian $H^*$ is set up as

\begin{equation}
\textbf{CM}_{2PN}(h)=\textbf{M}_{2PN}\otimes\textbf{A}_4(h).
\end{equation}

\begin{figure}
\includegraphics[width=10.5 cm]{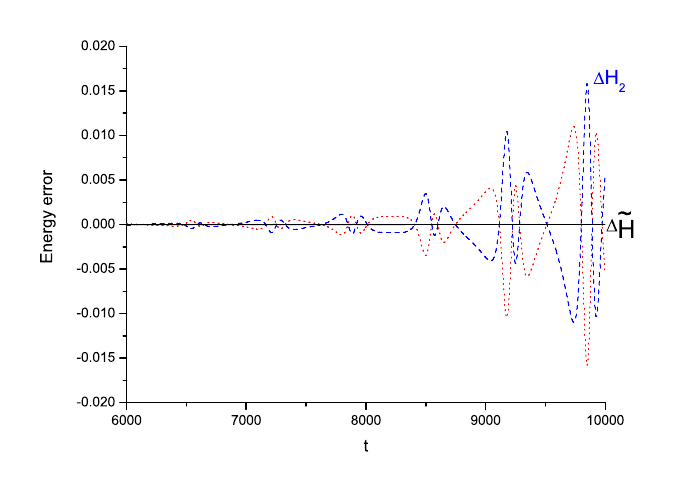}
\caption{{The 
 energy errors $\Delta E$ of the Hamiltonian in Equation \ref{eq:16}, as computed using the $\textbf{A}_4$ algorithm after the extended phase space, can be expressed as 
$\Delta\widetilde{H}=\widetilde{H}-2H(0)=\Delta H_1+\Delta H_2$, whereas $\Delta H_i=H_i(t)-H(0)$ and $H_i(t)$ represent the value of the Hamiltonian $H_i$ at time $t$. $H(0)$ denotes the initial value of the Hamiltonian in Equation \ref{eq:16}.} Time--axis symmetry exists between $\Delta H_1$(red dot) and $\Delta H_2$(blue dash).}\label{fig1}
\end{figure}

The $\textbf{A}_4$ algorithm {is treated as an explicit symplectic method serving to} the new Hamiltonian $\widetilde{H}$, ensuring effective preservation of the energy of $\widetilde{H}$, i.e., $\Delta \widetilde{H}=\Delta H_1+\Delta H_2\approx 0$. The error evolution of $H_1$ and $H_2$ shows a clear time--axis symmetry, as demonstrated in \mbox{Figure \ref{fig1}}. Specifically, if $H_1$ calculates more energy than the initial energy, $H2$ will calculate less, and vice versa. Taking advantage of this symmetry, we designed a manifold correction mapping approach to optimize the performance of the $\textbf{A}_4$ algorithm. The $\textbf{M}_{1PN}$ and $\textbf{M}_{2PN}$ corrections imposed on the solutions of $\textbf{A}_4$ serve three primary purposes. {The $\textbf{A}_4$ algorithm serves multiple purposes in relation to the new Hamiltonian $\widetilde{H}$. Firstly, it ensures that $H_1$ is equal to $H_2$ to prevent energy discrepancies that could impede the availability of numerical solutions. Secondly, it maintains the constancy of $\widetilde{H}$ after the correction, effectively suppressing the growth of energy errors. Thirdly, it reduces the energy deviation of each subterm of $H$ from half of the corresponding subterm of $\widetilde{H}$ through the correction process.
As the $\textbf{A}_4$ algorithm is an explicit symplectic method serving the new Hamiltonian $\widetilde{H}$, it accurately calculates the total energy as well as the energy of each individual subterm in $\widetilde{H}$; thus, the algorithm altering} these energies is not desirable, as it may weaken the algorithm's stability and precision.

In Section \ref{sec:4}, we will set initial values and perform numerical simulations of post-Newtonian Lagrangian and post-Newtonian Hamiltonian to compare the differences between the algorithms in terms of maintaining the constants of motion.

\section{Numerical Simulation}\label{sec:4}
\begin{figure}
\hspace{-9pt}\includegraphics[width=9.8 cm]{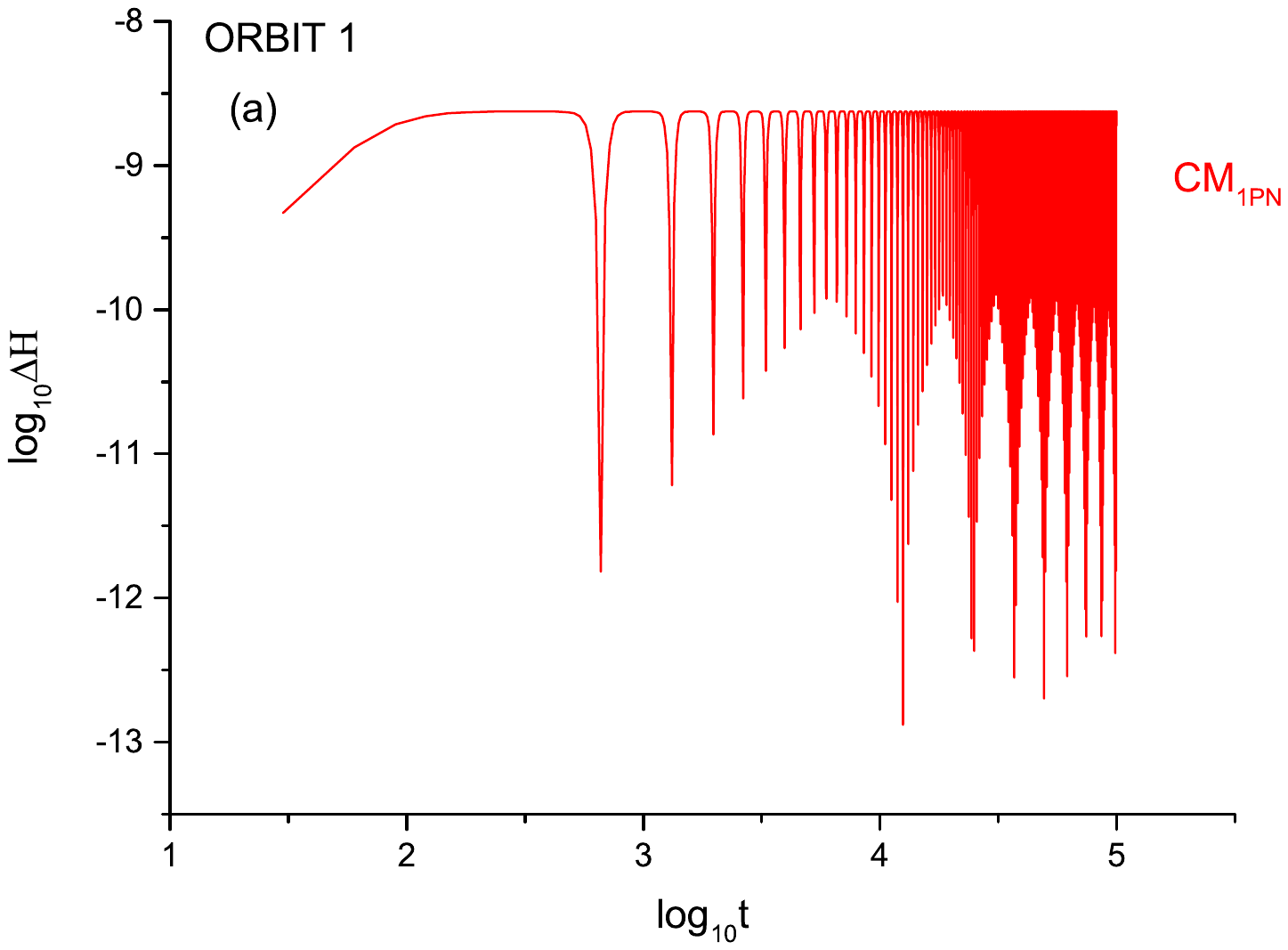}
\hspace{-9pt}\includegraphics[width=9.8 cm]{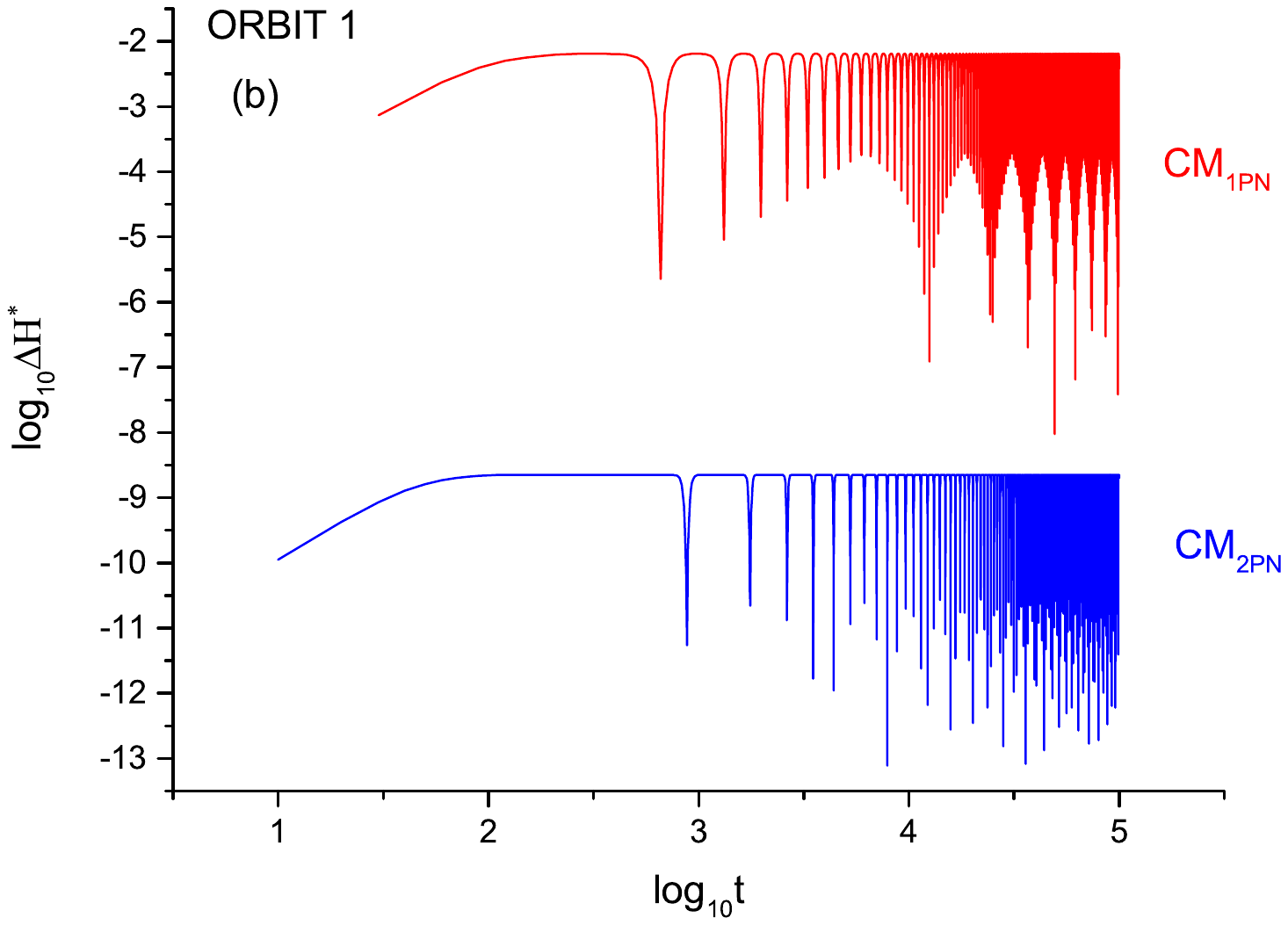}
\hspace{-9pt}\includegraphics[width=9.8 cm]{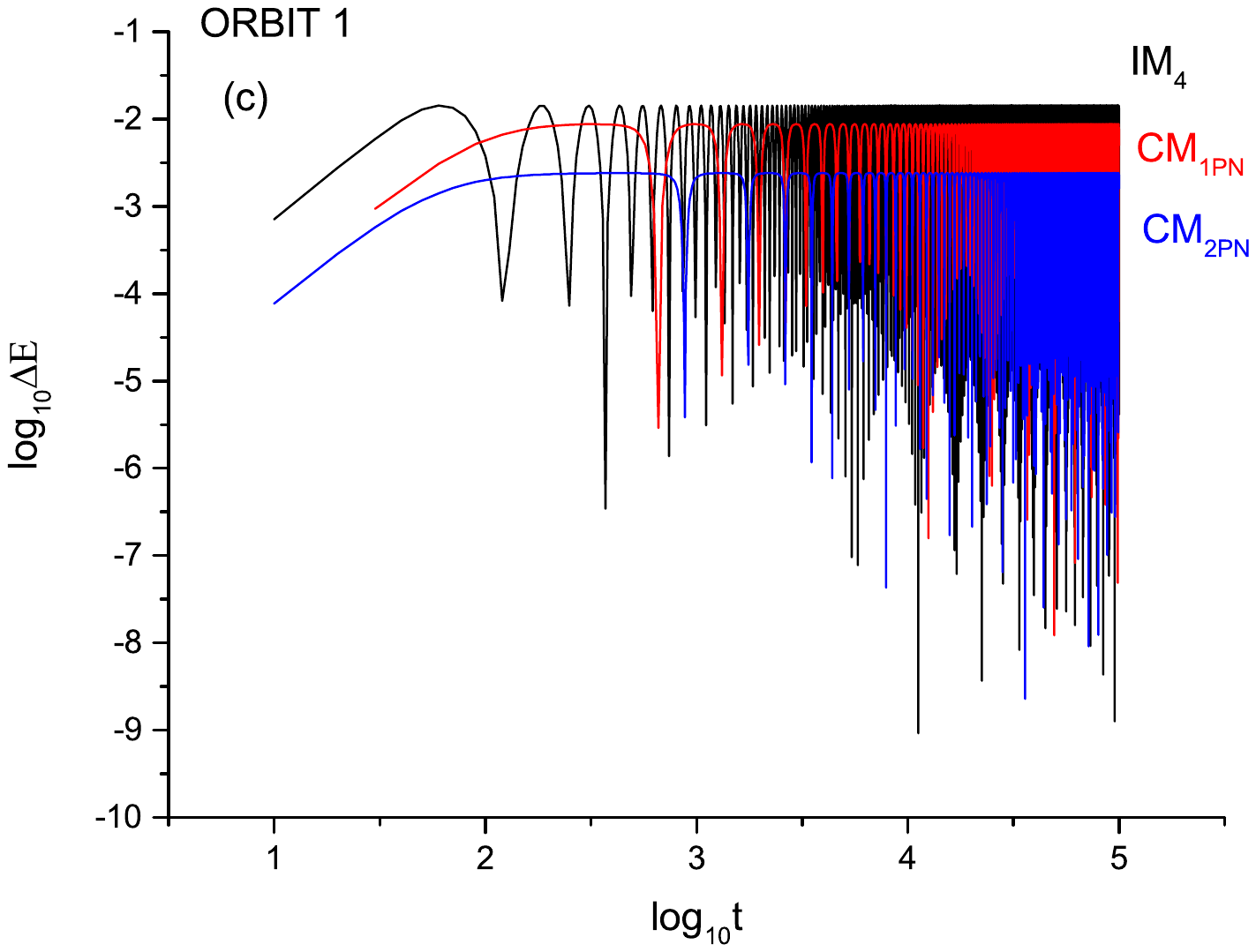}
\caption{Different 
 energy errors ($\Delta H, \Delta H^*, \Delta E$) in orbit 1.  { (\textbf{a}) The energy error of $H$, denoted as $\Delta H=|H(t)-H(0)|$, where $H(t)$ represents the value of the Hamiltonian $H$ at time $t$, and $H(0)$ is the initial value. (\textbf{b}) The energy error of $H^*$, denoted as $\Delta H^*=|H^*(t)-H^*(0)|$. (\textbf{c}) The energy error of $E$, denoted as $\Delta E^*=|E(t)-E(0)|$. The algorithm $\textbf{IM}_4$ is drawn with a black line, whereas $\textbf{CM}_{1PN}$ and $\textbf{CM}_{2PN}$ are drawn with red and blue lines, respectively}.} \label{fig2}
\end{figure}

This section showcases the outcomes of our numerical simulations, where we compare the post-Newtonian Lagrangian and post-Newtonian Hamiltonian algorithms in maintaining the constants of motion. To this end, we set initial values for a specific orbit, named orbit 1, with initial conditions $(\beta;\textbf{r},\textbf{v})=(\frac{5}{4};10,0,0,0,0.52,0)$. The initial value of the momentum $\textbf{p}$ in the post-Newtonian Hamiltonian is obtained from Equation \eqref{eq:11}. We use the fourth-order implicit midpoint method ($IM_4$) to calculate the 1PN Lagrangian, whereas the algorithms $\textbf{CM}_{1PN}$ and $\textbf{CM}_{2PN}$ are used to calculate the Hamiltonian $H$ and $H^*$, respectively. We take a fixed step size of $h=1$ and plot the energy errors in Figure \ref{fig2}a,b. We observe that the $\textbf{CM}_{1PN}$ algorithm designed for the Hamiltonian $H$ has significantly better accuracy in terms of energy error compared to $IM_4$. However, the accuracy of $\textbf{CM}_{1PN}$ drops considerably in the $H^*$ error behavior, as expected due to the vanishing of the 2PN term, whereas the error in $\textbf{CM}_{2PN}$ is at an order of $10^{-8}$. Finally, Figure \ref{fig2}c shows that $\textbf{CM}_{2PN}$ performs the best in terms of accuracy and long-term stability compared to $\textbf{CM}_{1PN}$ and $IM_4$.

\begin{figure}
\includegraphics[width=10.5 cm]{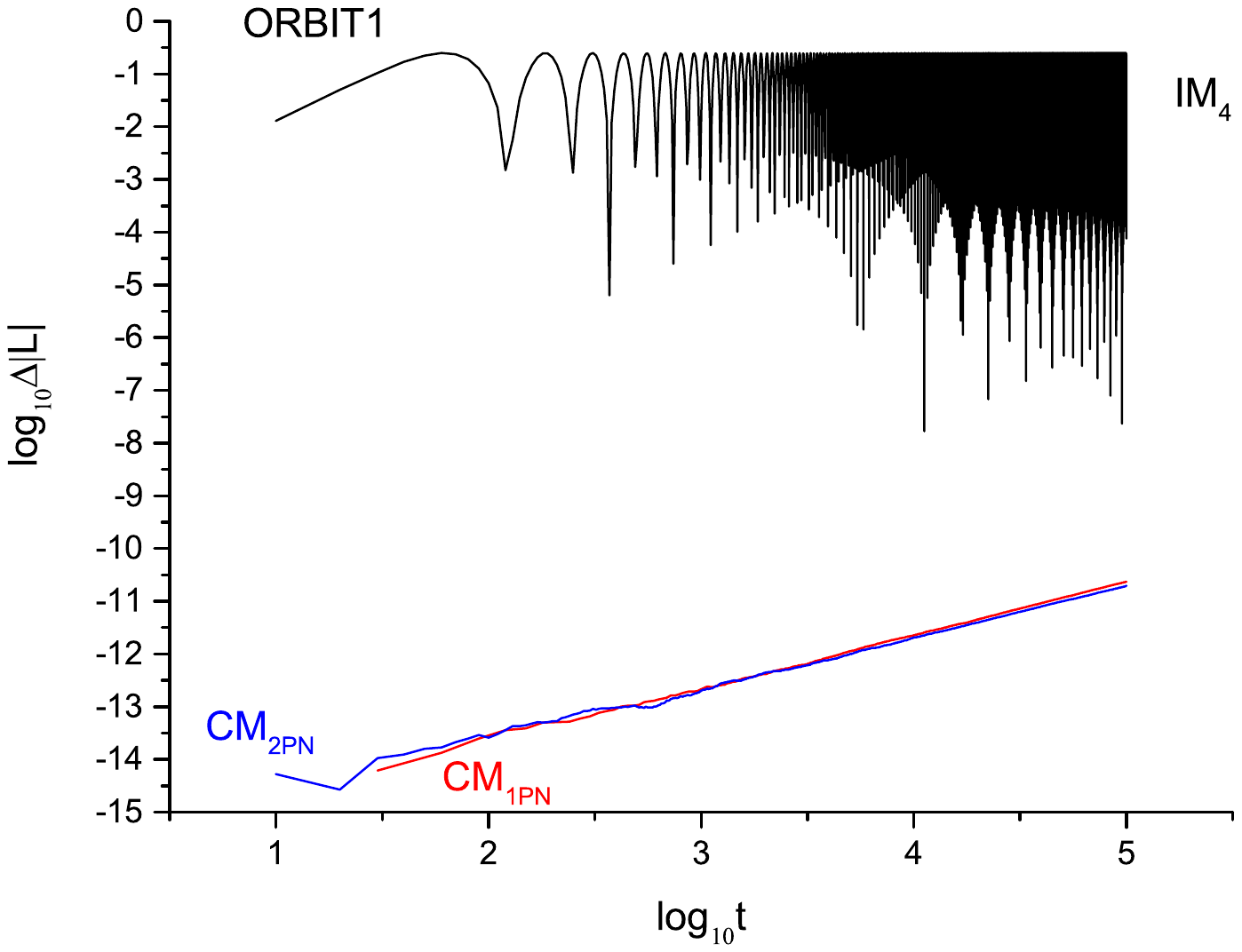}
\caption{{The 
 angular momentum errors in orbit 2, denoted as $\Delta \L=\L-\L_0$, where $\L=|\textbf{\L}|$ and $\textbf{\L}$ is calculated using three different methods: $IM_4$ (represented by black), $\textbf{CM}_{1PN}$ (represented by red), and $\textbf{CM}_{2PN}$ (represented by blue).}\label{fig3}}
\end{figure}

 Aside from ensuring energy conservation, we also track the preservation of orbital angular momentum $\textbf{\L}=\mathbf{r}\times\mathbf{p}=$$[1+\frac{1}{c^2}(\frac{1-3\eta}{2}v^2+\frac{3+\eta}{r})]\mathbf{r}\times\mathbf{v}$, and examine its error as another performance metric for the algorithms.  {Figure \ref{fig3} depicts the angular momentum errors $\Delta \L=\L-\L_0$ with $\L=|\textbf{\L}|$, where $\L_0$ represents the initial value.} We deduce that the performance of the $IM_4$ algorithm in terms of angular momentum error is similar to its performance in energy error, with the worst accuracy but good long-term stability.  {Conversely, both $\textbf{CM}_{1PN}$ and $\textbf{CM}_{2PN}$ exhibit exceptional performance in terms of angular momentum error}, with very little difference between them and significantly superior to $IM_4$. However, there is a noticeable error growth in $\textbf{CM}_{1PN}$ and $\textbf{CM}_{2PN}$. Our simulations show that the algorithm $\textbf{CM}_{1PN}$ in the 1PN Hamiltonian has a significant advantage over the $IM_4$ algorithm in maintaining the conservation of orbital angular momentum and a small accuracy advantage in maintaining energy integrals. The algorithm $\textbf{CM}_{2PN}$ in the 2PN Hamiltonian also has a significant advantage in maintaining angular momentum, while being comparable to $\textbf{CM}_{1PN}$, with some improvement in the accuracy of the energy.

\vspace{-6pt} 

\begin{figure}
\includegraphics[width=8.2 cm]{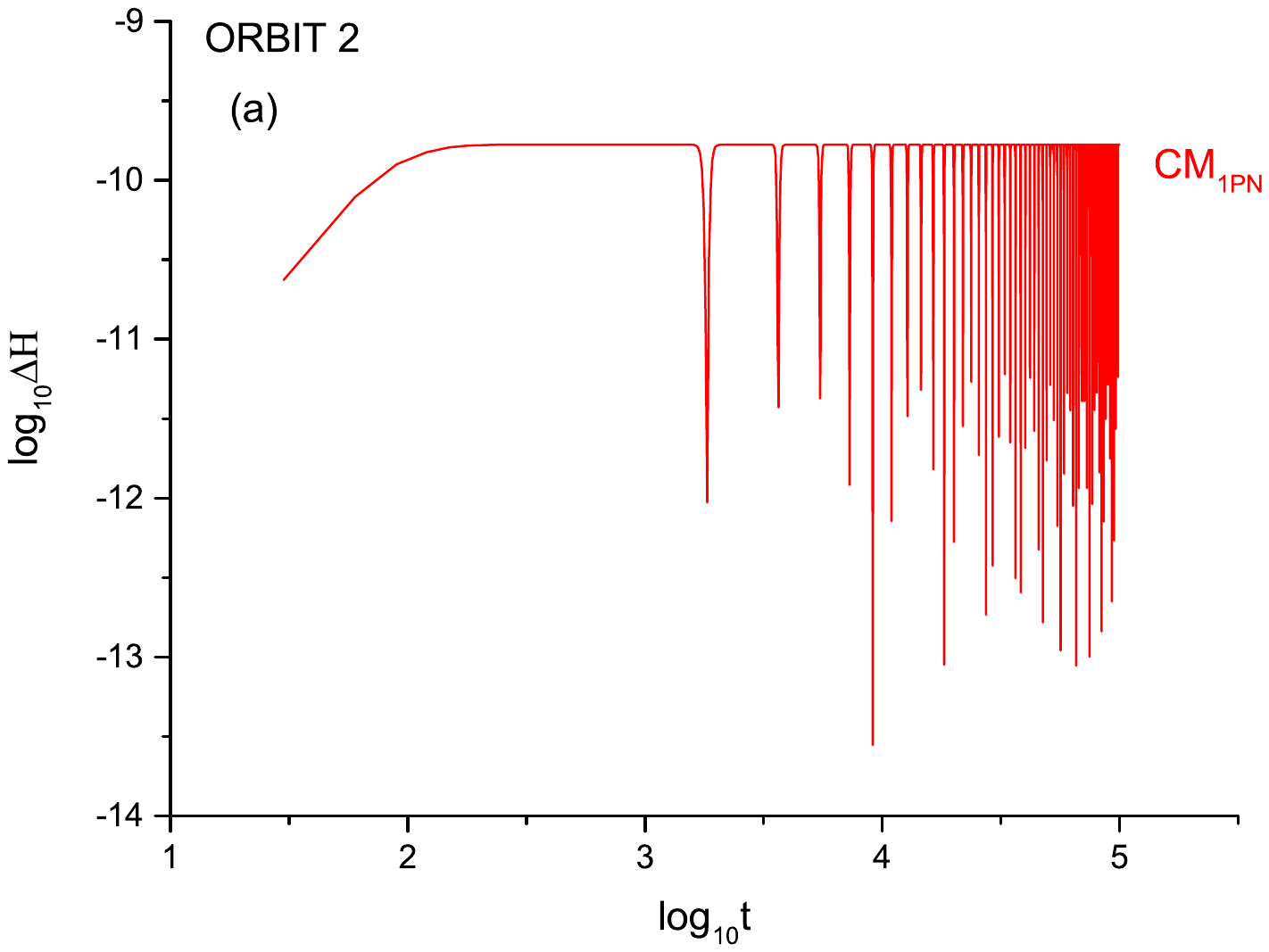}\\
\includegraphics[width=8.2 cm]{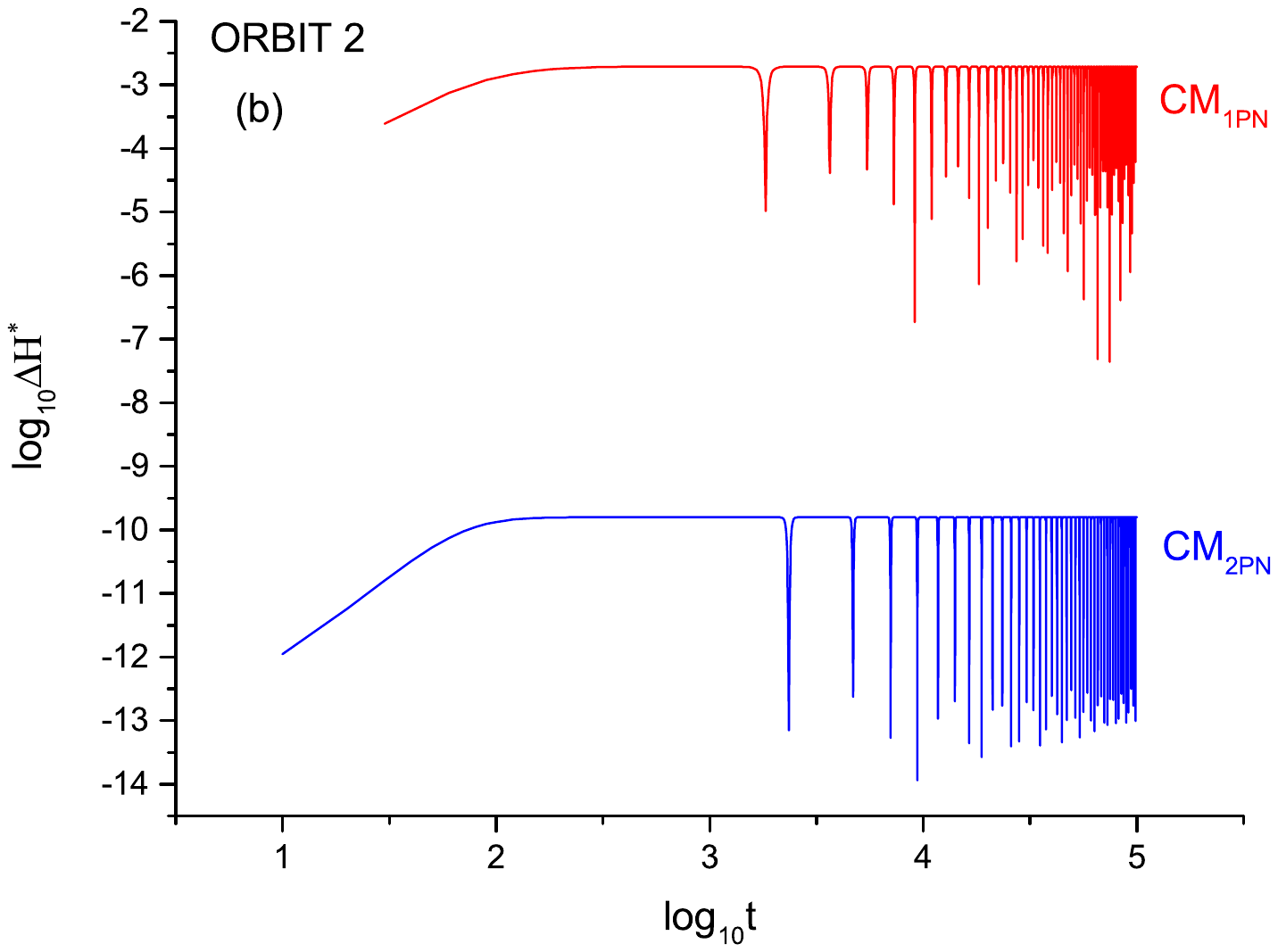}\\
\includegraphics[width=8.2 cm]{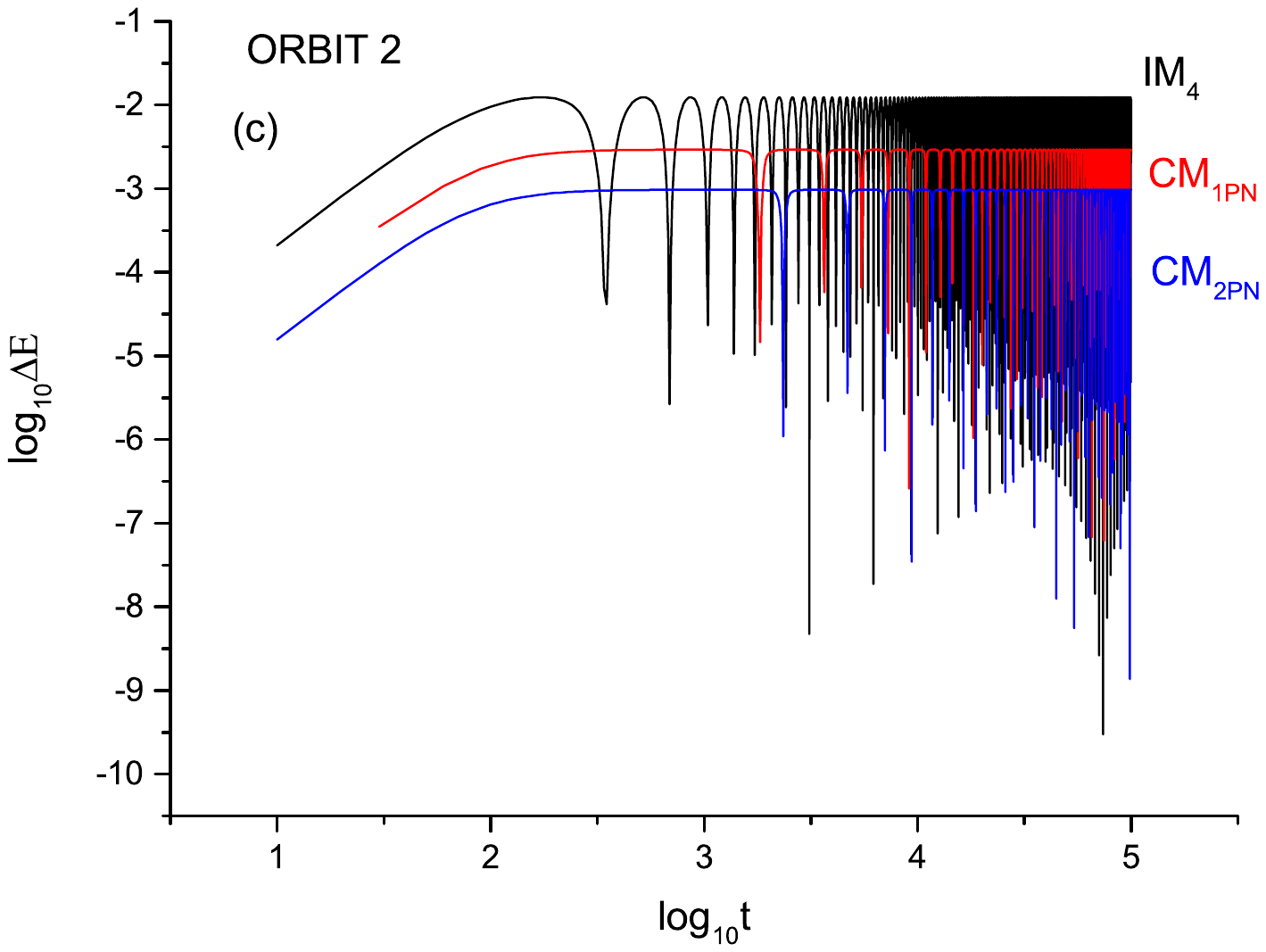}
\caption{Different 
 energy errors($\Delta H, \Delta H^*, \Delta E$) in orbit 2.  {(\textbf{a}) The energy error of $H$, represented as $\Delta H$, is calculated as the absolute difference between the value of the Hamiltonian $H$ at time $t$ ($H(t)$) and its initial value ($H(0)$). (\textbf{b}) The energy error of $H^*$, denoted as $\Delta H^*$, is determined as the absolute difference between the value of the Hamiltonian $H^*$ at time $t$ ($H^*(t)$) and its initial value ($H^*(0)$). \mbox{(\textbf{c}) The energy} error of $E$, denoted as $\Delta E^*$, is computed as the absolute difference between the value of $E$ at time $t$ ($E(t)$) and its initial value ($E(0)$). The algorithm IM4 is represented by a solid black line, whereas C1PN and C2PN are indicated by dashed red and blue lines, respectively. The performance of each algorithm in orbit 2 is similar to that of orbit 1}.\label{fig4}}
\end{figure}

 {To validate the aforementioned conclusion, additional numerical simulations will be conducted in a different orbit, referred to as orbit 2. The initial conditions for orbit 2 are set as $(\beta;\textbf{r},\textbf{v})=(\frac{5}{4};10,0,0,0,0.52,0)$. In Figure \ref{fig4}, three categories of energy error in orbit 2 will be depicted as follows:
(a) Energy error analysis of orbit 2 for $IM_4$, $\textbf{CM}_{1PN}$, and $\textbf{CM}_{2PN}$ will be presented in Figure \ref{fig4}a.
(b) Figure \ref{fig4}b will display the energy error analysis of orbit 2 specifically for $IM_4$.
(c) The energy error analysis of orbit 2, focusing on $IM_4$, will be illustrated in Figure \ref{fig4}c.}

 {It is evident from the figures that the performance of $IM_4$, $\textbf{CM}_{1PN}$, and $\textbf{CM}_{2PN}$ in orbit 2 closely resembles that of orbit 1. $IM_4$ continues to exhibit the highest error, $\textbf{CM}_{1PN}$ demonstrates a widening gap with $IM_4$, and $\textbf{CM}_{2PN}$ remains the most advanced in terms of accuracy.}

\begin{figure}
\includegraphics[width=10.5 cm]{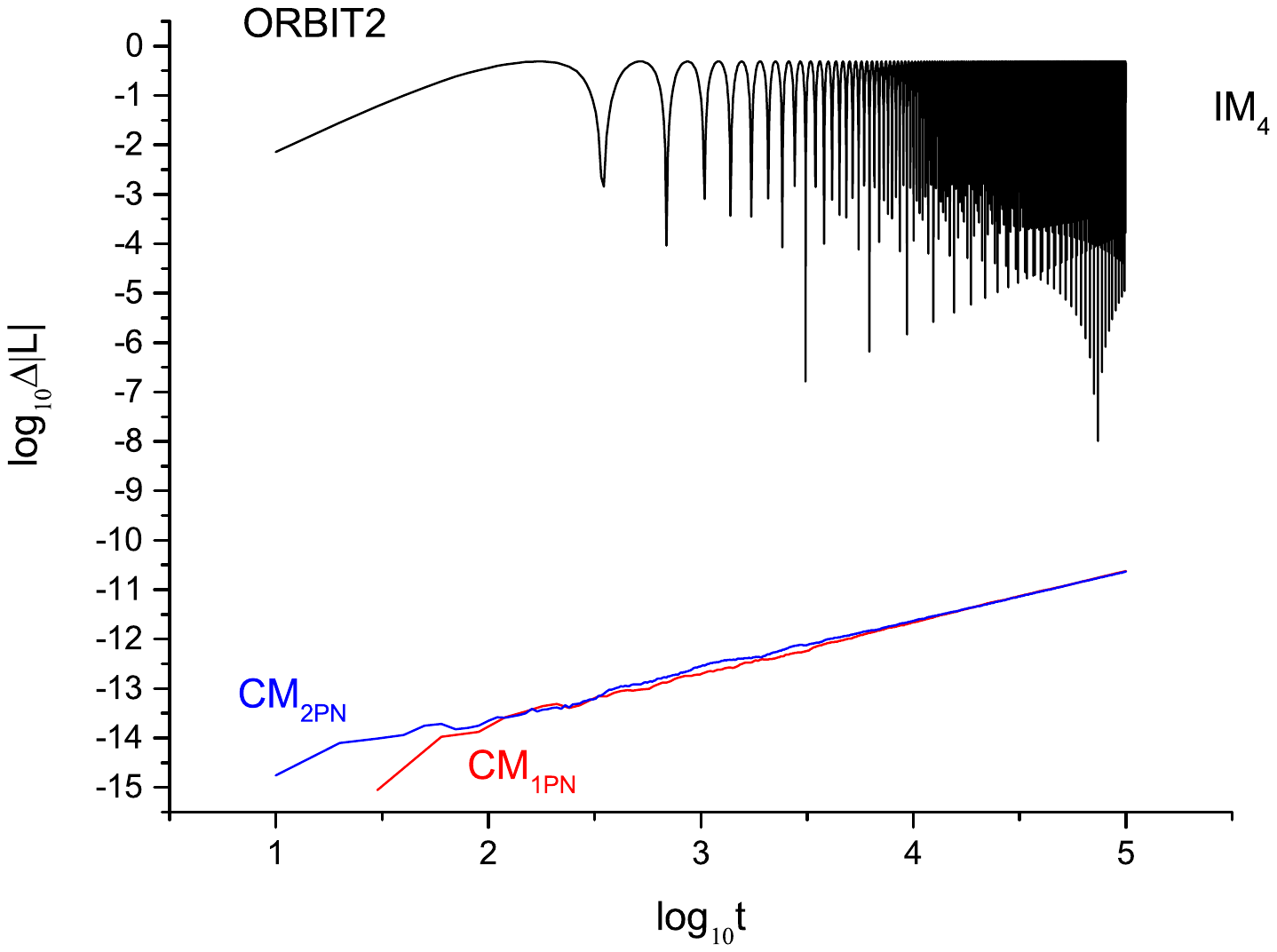}
\caption{The 
 errors of the angular momentum $\Delta \L=\L-\L_0$ in orbit 2, here $\L=|\textbf{\L}|$, $\textbf{\L}$ are calculated by $IM_4$ (black), $\textbf{CM}_{1PN}$ (red), $\textbf{CM}_{2PN}$ (blue).\label{fig5}}
\end{figure}

 {Turning to the angular momentum errors depicted in Figure \ref{fig5}, it is observed that there is no significant improvement for $IM_4$, which still exhibits considerable deviation compared to the first-order post-Newtonian approximation, $\textbf{CM}_{1PN}$. Furthermore, the inclusion of the second-order post-Newtonian term in $\textbf{CM}_{2PN}$ does not contribute significantly to reducing the angular momentum error.}

 {Summarizing the findings from the numerical simulations conducted for both orbit 1 and orbit 2, we can conclude that $\textbf{CM}_{1PN}$ and $\textbf{CM}_{2PN}$ perform better for the calculation of the post-Newtonian approximation to the Hamiltonian. They exhibit a slight advantage in terms of energy error while demonstrating a notably superior accuracy in the calculation of angular momentum.}
\vspace{-6pt} 
\section{Summary}\label{sec:5}
The exact equations of motion for a post-Newtonian Lagrangian formalism are the Euler--Lagrange equations, which consist of a coherent Lagrangian without any truncated terms. However, when the post-Newtonian Lagrangian form of general relativity maintains only a certain post-Newtonian order, it is referred to as the incoherent Lagrangian, with higher-order terms of the acceleration truncated. Incoherent Lagrangian can be numerically simulated using the Runge--Kutta method and implicit algorithms. Therefore, in the incoherent Lagrangian, motion constants such as energy integrals are only approximately conserved. The retention of the Hamiltonian in a certain post-Newtonian (PN) order leads to a high-order truncation problem. In addition, if the Hamiltonian is separable, symplectic algorithms can be used, which provide excellent performance. For the case where the Hamiltonian is inseparable, symplectic-like algorithms such as the  {phase-space expansion method with correction map, namely correction map method,} can be used. The  {phase-space expansion} method with correction map is referred to as the correction map method, which utilizes the symmetry of energy errors in $H_1$ and $H_2$ in the new Hamiltonian $\widetilde{H}$ to improve the accuracy and stability of the algorithm. 

A comparison was made between the performance of the implicit midpoint method in the incoherent Lagrangian and the correction map method in the PN Hamiltonian. Under the 1PN Newtonian approximation, the correction map method performed better in terms of energy error, exhibiting higher accuracy and comparable stability. On the other hand, with regards to angular momentum error, the correction map method was significantly higher, reaching an order of $10^{-11}$, whereas the implicit midpoint method was only $10^{-1}$. Similarly, under the 2PN Newtonian Hamiltonian, the manifold correction mapping method further improved the accuracy of energy error, but there was no noticeable impact on the angular momentum error.

In conclusion,  {we compared the implicit midpoint methods for solving the equations of motion in post-Newtonian Lagrangians and the correction map method for PN Hamiltonians and investigate the extent to which both methods can uphold invariance of the motion's constants, such as energy conservation and angular momentum preservation. Ultimately, the results of numerical simulations demonstrate the superior performance of the correction map method, particularly with respect to angular momentum conservation.} Compared to incoherent Lagrangian, we recommend using the manifold correction map method for the Hamiltonian of compact binaries as a numerical tool.

\end{document}